\documentclass[a4paper,11pt]{article}

\usepackage{textcomp}
\usepackage{amssymb}
\usepackage{units}

\usepackage[overload]{textcase}

\usepackage{graphicx}
\usepackage{subfigure}
\usepackage[labelsep=period]{caption}

\usepackage{amssymb}

\usepackage{amsmath}
\usepackage{mathabx}

\usepackage[sort&compress]{natbib}
\bibpunct{}{}{,}{s}{}{}

\begin{document}

\thispagestyle{plain}
\setcounter{page}{1}

\vfill

\begin{center}

   \vspace{0.3cm}
   {\Large \textbf{Uptake of water droplets by nonwetting capillaries}} \\
   \vspace{0.3cm}
   {\large Geoff R. Willmott$^{\dag 1}$, Chiara Neto$^{2}$ and Shaun C. Hendy$^{1,3}$}\\
   {\large \emph{$^1$The MacDiarmid Institute for Advanced Materials and Nanotechnology, Industrial Research Limited, 69 Gracefield Rd, PO Box 31-310, Lower Hutt 5040, New Zealand}}\\
   {\large \emph{$^2$School of Chemistry, The University of Sydney, NSW2006, Australia}}\\   
   {\large \emph{$^3$School of Chemical and Physical Sciences, Victoria University of Wellington, Wellington 6140, New Zealand}}\\
	\vspace{0.3cm}
	{\large $^\dag$Corresponding author}\\
	\vspace{0.3cm}
	{\large Email: g.willmott@irl.cri.nz}\\
	\vspace{0.3cm}
    {\large Phone: (64) (0)4 931 3220}\\
    \vspace{0.3cm}
    {\large Fax: (64) (0)4 931 3117}\\
   \vspace{0.3cm}

\begin{abstract}
\noindent 
We present direct experimental evidence that water droplets can spontaneously penetrate non-wetting capillaries, driven by the action of Laplace pressure due to high droplet curvature. Using high-speed optical imaging, microcapillaries of radius 50 to 150~$\mu$m, and water microdroplets of average radius between 100 and 1900~$\mu$m, we demonstrate that there is a critical droplet radius below which water droplets can be taken up by hydrophobised glass and polytetrafluoroethylene (PTFE) capillaries. The rate of capillary uptake is shown to depend strongly on droplet size, with smaller droplets being absorbed more quickly. Droplet size is also shown to influence meniscus motion in a pre-filled non-wetting capillary, and quantitative measurements of this effect result in a derived water-PTFE static contact angle between 96\textdegree\space and 114\textdegree. Our measurements confirm recent theoretical predictions and simulations for metal nanodroplets penetrating carbon nanotubes (CNTs). The results are relevant to a wide range of technological applications, such as microfluidic devices, ink-jet printing, and the penetration of fluids in porous materials. 
\end{abstract}
\end{center}

\clearpage

\section{Introduction}

The penetration of a liquid into a capillary tube, driven by surface forces, has been a topic of active research for at least the last century, and is important in a wide range of fields. Examples of processes utilizing capillarity include transport of liquids in plants and imbibition of fluids by porous media such as powders, soils and granular materials. Studies of capillary uptake have been applied to oil extraction, textiles, paper and printing, and in weightless environments. Nanoscale capillary research has commenced with molecular dynamics simulations used to study fluid uptake by carbon nanotubes (CNTs) \cite{823,757,758} and a wetting gold nanopore.\cite{816} High speed photography has been used to study nanoporous silica \cite{815} and rectangular nanochannels.\cite{832} Capillary techniques are also fundamental in the growing field of microfluidics, and some research on capillary uptake has been carried out in that context.\cite{805,818} 

\begin{figure}
\begin{center}
\subfigure[]{\label{Fig1aa}\includegraphics[width=4.5cm]{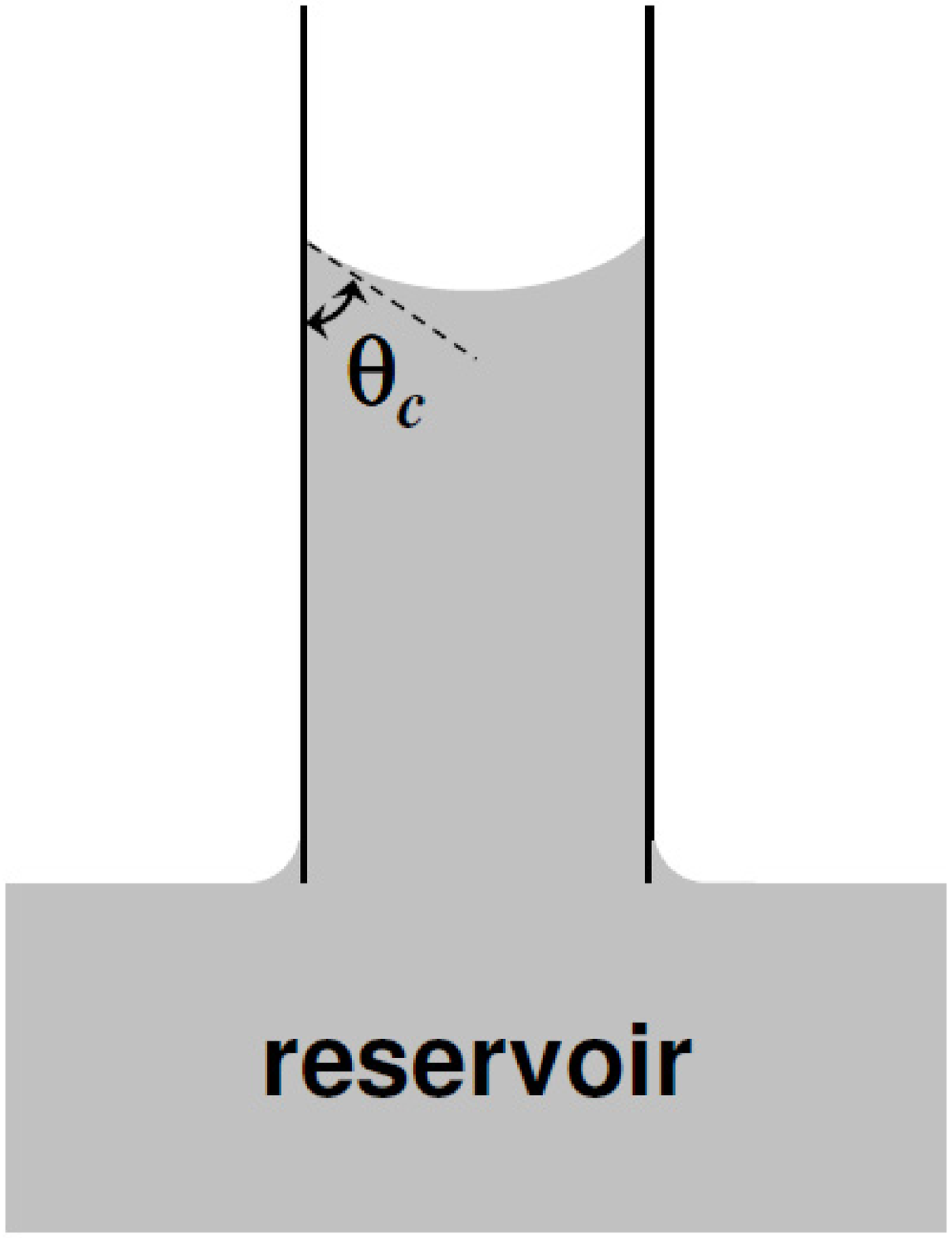}}
\subfigure[]{\label{Fig1a}\includegraphics[width=3.15cm]{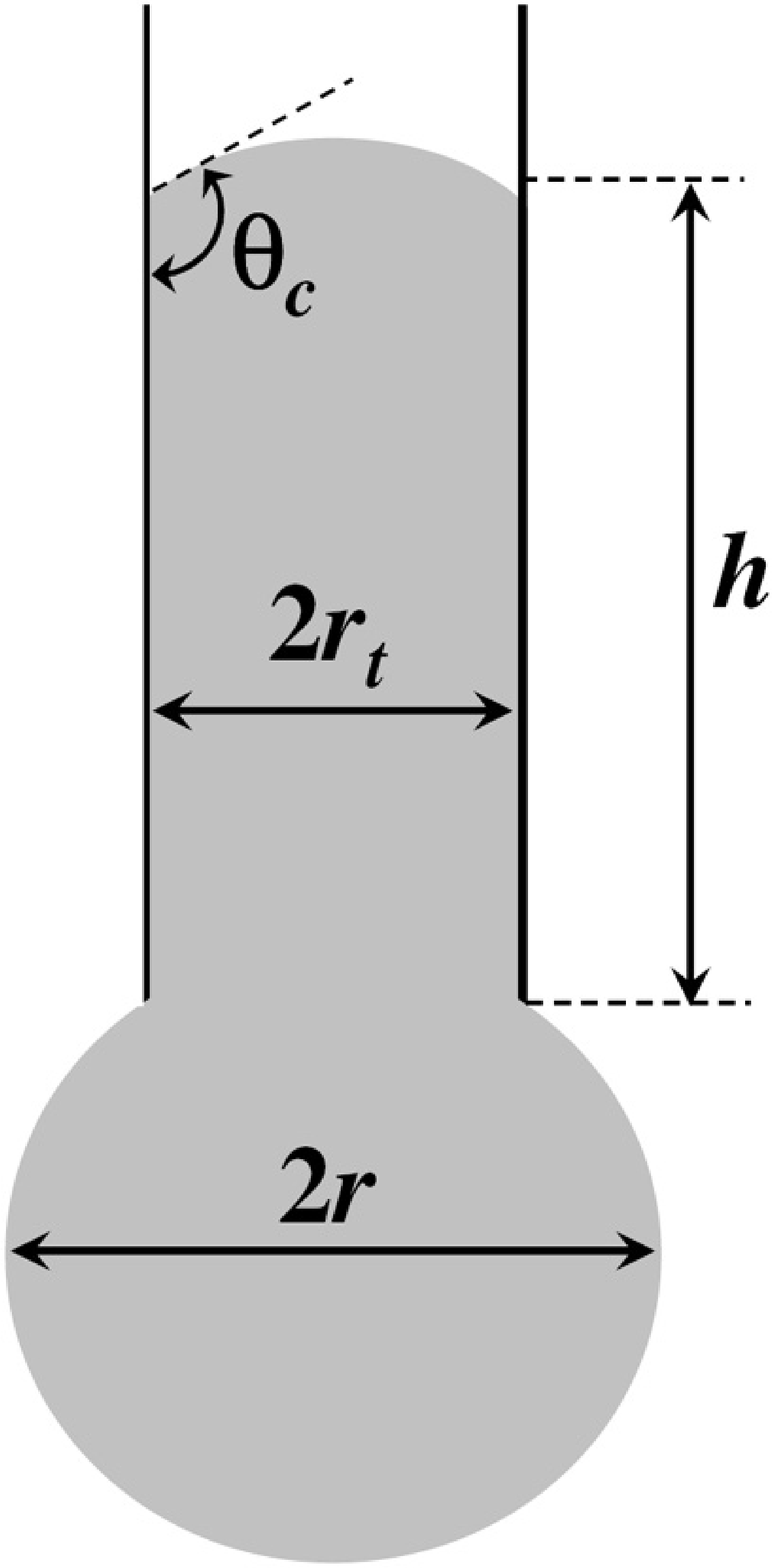}}
\subfigure[]{\label{Fig1b}\includegraphics[width=8.25cm]{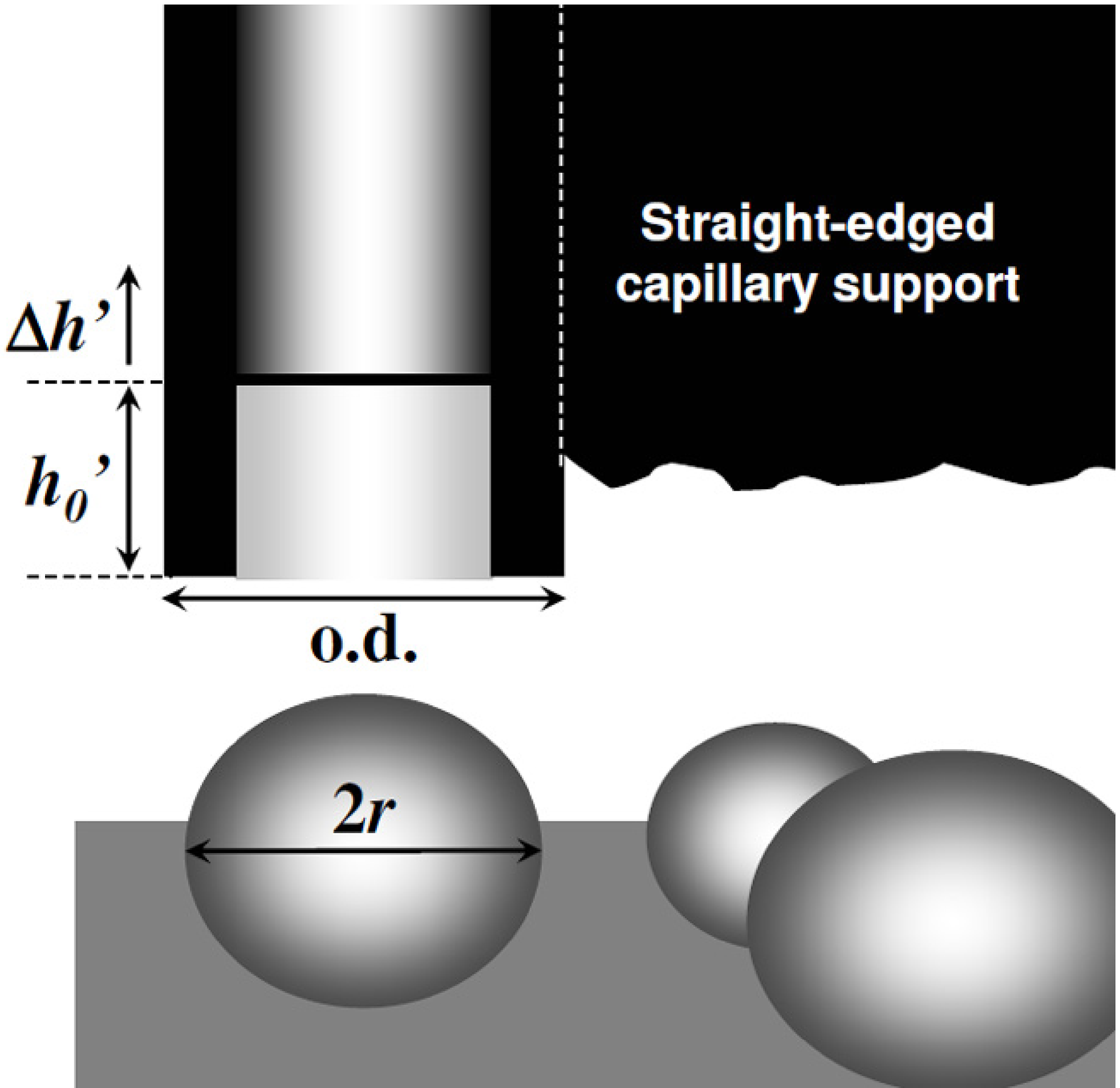}}
\end{center}
\caption{Fig.~\ref{Fig1aa} shows usual capillary uptake by a wetting capillary. Fig.~\ref{Fig1a} defines the geometry of the model system discussed in the text. The spherical-cap fluid droplet has dynamic viscosity $\eta$ and an internal Laplace pressure caused by surface tension $\gamma$ at the fluid-air boundary. Fig.~\ref{Fig1b} schematically shows what is observed in an image from a typical experiment. The capillary is shown partially filled, with the near-horizontal meniscus visible at the centre of the tube. The meniscus height measurement prior to the tube making contact with the droplet is $h'_0$, and $\Delta h'$ is the change in height thereafter. Images are spatially calibrated using the known outer diameter of the capillary. The straight-edged capillary support is used to prevent the PTFE tubes from bending. Several drops on a superhydrophobic surface may be observed in any photo, and one drop is aligned with the base of the capillary tube. The superhydrophobic surface is raised and lowered to investigate the drop-capillary interaction.}
\label{Fig1}     
\end{figure}

The simplest capillary uptake configuration, in which a macroscopic tube is immersed in a liquid reservoir, is well understood:\cite{842} when the capillary tip contacts the liquid-air interface of the reservoir, a meniscus of the liquid forms inside the capillary. If the liquid wets the capillary walls, i.e. its contact angle with the wall ($\theta_c$, Fig.~\ref{Fig1aa}) is less than 90\textdegree, the liquid meniscus takes on a concave curvature. Due to surface tension and this curvature, there is a pressure difference (Laplace Pressure) across the meniscus which drives the liquid into the capillary. The meniscus stops rising when an equal and opposing pressure is applied, for example due to gravity. In this classical macroscopic picture, liquids with contact angles higher than 90\textdegree\space cannot rise in the tube by capillary action, and instead will form a meniscus with convex curvature at the base of the tube.

The present paper is concerned with a capillary coming into contact with an isolated liquid droplet of similar dimensions to the tube (Fig.~\ref{Fig1a}). In this case, the effect of the Laplace pressure on the droplet itself needs to be considered, as highlighted by Marmur \cite{790} and Schebarchov and Hendy.\cite{757,758} We use physical experiments to test the theoretical predictions of these workers for the first time, and in particular confirm three key effects:
\begin{enumerate}
\item The finite size of a droplet, and its consequent high Laplace pressure, allows capillary absorption to occur, even in non-wetting capillary tubes.
\item The direction of meniscus motion within a non-wetting capillary depends on droplet size.
\item The speed of uptake of a droplet depends on droplet size.
\end{enumerate}

Despite the importance of these predictions, they remain largely untested by experiments. Marmur's analysis was extended to the radial capillary \cite{789,794} and displacement of fluids in a tube,\cite{789,796} but was mostly applied to porous membranes (e.g. paper) for which drop size is large relative to effective pore size.\cite{797} To the best of our knowledge, Marmur's work has been mentioned in one previous study of droplet capillary uptake,\cite{807} in which data was obtained by measuring the volume of the protruding droplet as a function of time. The data presented in that study showed a small decrease of the capillary wetting rate with increasing water drop size for a 0.3~mm radius glass capillary. Some previous experiments have employed chemical treatments to make capillary surfaces more hydrophobic,\cite{828,829} but presumably still wetting ($\theta_c<$~90\textdegree ). Ichikawa and Satoda's experiments \cite{834} included investigation of receding capillary motion for water in a horizontal perfluoroalkoxy tube, for which $\theta_c>$~90\textdegree. Typically, experimental studies of single capillary uptake dynamics have used high-speed photography,\cite{807, 809,810,828,829,831,833,834} reservoirs of strongly wetting fluids ($\theta_c=0$) and capillaries between 0.1 and 1~mm diameter. The use of relatively large tubes has been enabled by limited-gravity experiments,\cite{813,824} while high viscosity fluids (e.g. silicone oils) have been used to slow the dynamics in some cases.\cite{809,831}  

Theoretical development has concentrated on a model system consisting of a single, cylindrical tube in contact with an incompressible, Newtonian fluid at low Reynolds number. The governing differential equation for the height of fluid $h$ in a vertical capillary in contact with a spherical drop of water of radius $r$ can be written

\begin{equation}\label{eq:812}
\rho\frac{d}{dt}\left(h\frac{dh}{dt}\right) + \frac{8\eta}{r_t^2}h\frac{dh}{dt}=\frac{2\gamma\cos\theta_c}{r_t} - \rho gh + \frac{2\gamma}{r}.   
\end{equation}

\noindent Here $\rho$ is fluid density, $\eta$ and $\gamma$ are the dynamic viscosity and surface tension of the fluid, $t$ is time, $g$ is gravitational acceleration and $r_t$ is the radius of the cylindrical capillary. Equation~\ref{eq:812} represents a balance between the following force terms (from left to right): fluid inertia, viscous drag, capillary force, gravity and Laplace pressure in the droplet. Theoretical descriptions of capillary uptake have often discussed Eq.~\ref{eq:812} in the limit as $r\rightarrow\infty$.\cite{809,812,814,817,821,820,824,826,828,834} The last term, which has not been widely studied, accounts for the Laplace pressure within a spherical drop in contact with the capillary (Fig.~\ref{Fig1a}), and is large for small droplets. 

With the inclusion of the Laplace pressure term, the interesting predictions described above are clearly illustrated. For a single capillary in contact with a small droplet, penetration should occur if the right hand side of Eq.~\ref{eq:812} 
is positive (as long as $d^2h$~/~$dt^2$ is not negative). It is therefore possible for a dry capillary to take up a fluid when $\theta_c > 90$\textdegree, as long as the drop radius is smaller than a threshold maximum value,

\begin{equation}\label{eq:rmax}
r_{max} = -\frac{r_t}{\cos\theta_c} .
\end{equation}

\noindent The main quantitative focus of the present work is experimentally probing this threshold. The size of the drop should additionally affect the direction and rate of meniscus motion, and the hydrostatic limit at high Bond number. 

Even for a wetting tube in contact with a fluid reservoir ($r=\infty$), the uptake dynamics depend strongly on the experimental conditions. In this work, we are concerned with experiments at low Bond number\cite{824} Bo~$= \left(\rho g r_t^2\right)$ / $\gamma$, for which gravity can be neglected. Even so, three regimes have been identified for the reservoir case, in which different terms of Eq.~\ref{eq:812} dominate: 
\begin{enumerate}
\item `Entrance effects' regime: when the fluid first enters the tube, inertia and the capillary forces dominate, because the viscous term is zero at $h=0$. In this regime, meniscus reorientation and pre-linear inertial acceleration (not considered in Eq.~\ref{eq:812}) are important, and analyses suggest that $h$~$\propto$~$t^2$.\cite{809, 824, 837}
\item `Bosanquet' regime: when entrance effects are no longer important, but inertia and the capillary forces still dominate. The first and third terms of Eq.~\ref{eq:812} can be used to find a singular solution to this problem (attributed to Bosanquet\cite{838}) at $t=0$, for which $h$~$\propto$~$t$.\cite{809,812,834} 
\item `Lucas-Washburn' regime: in which the viscous force dominates the inertial term. The well-known analytic expression for this regime gives a solution for which $h$~$\propto$~$t^\frac{1}{2}$, or in full:\cite{791,792} 

\begin{equation}\label{eq:L&W}
h\left(t\right)=\left(\frac{\gamma r_t \cos\theta_c}{2\eta}\right)^{\frac{1}{2}}\sqrt{t}.
\end{equation}
\end{enumerate}

When a droplet is in contact with the capillary, the last term in Eq.~\ref{eq:812} (the Laplace term) is significant. This term was first postulated by Marmur \cite{790}, who used an approximation to solve Eq.~\ref{eq:812} excluding only inertia. Schebarchov and Hendy \cite{757,758} provided an analytical solution to the equation 

\begin{equation}\label{eq:S&H}
\frac{8\eta}{\left(2b+r_t\right)r_t}\frac{dh}{dt}=\frac{2\gamma}{h}\left(\frac{1}{r}+\frac{\cos\theta_c}{r_t}\right),
\end{equation}

\noindent which is Eq.~\ref{eq:812} excluding inertia and gravity, and incorporating non-zero slip length $b$. Equation~\ref{eq:S&H} reduces to the Lucas-Washburn regime for a reservoir ($r=\infty$) and $b=0$. Equation~\ref{eq:S&H} also enables us to study the slip length $b$, a topic of much recent interest,\cite{303,724} using capillary uptake dynamics. To our knowledge, no theoretical work has addressed the Laplace term for the droplet under conditions equivalent to the `Entrance effects' and `Bosanquet' regimes. In this paper, we will clarify that there are further assumptions of this model to be addressed, even for a well-defined experimental system. We therefore concentrate our quantitative work on the threshold $r_{max}$ (Eq.~\ref{eq:rmax}), which is key to explaining our new interesting observations and common to both inertia- and viscous-dominated regimes.

In further theoretical work, boundaries between the three regimes have been discussed,\cite{817,824} analytic solutions to Eq.~\ref{eq:812} have been obtained under various specific conditions,\cite{820,821,826} and approaches to dimensionless scaling have been described.\cite{814} Other studies have addressed uptake of viscoelastic fluids \cite{807,808} and accounted for the reservoir container geometry in microgravity.\cite{824} The importance of the dynamic contact angle has emerged,\cite{810,821,824,825,826} while numerical, molecular dynamics and statistical dynamics studies have also been carried out.\cite{757,758,823,816,819,822,825,830} 

We have carried out experiments in conditions which cover all three of the uptake regimes, as defined for the equivalent reservoir uptake experiments.\cite{824,809} Direct comparison of regimes is difficult, because our experiments use capillaries with a high contact angle. Additionally, capillaries are sometimes pre-filled, droplet geometry can rapidly change during uptake and both filling and draining are observed. We expect that these demonstrative experiments will open up a range of opportunities for further theoretical and experimental development. For example, transitions between regimes \cite{809,829,824} and the effect of the dynamic contact angle \cite{810,829,831} might be further investigated.

\section{Experimental}

\subsection{Materials}

Droplets of deionised water (18.2~M$\Omega$~cm) were used in experiments with silanized glass and polytetrafluoroethylene (PTFE) capillary tubes (see Table~\ref{Tab} for details). As-supplied tubing was cut using tungsten carbide and stainless steel blades for glass and PTFE respectively. Capillaries were inspected under a microscope to ensure that the ends were reasonably smooth and azimuthally symmetric, although not perfectly so. 

Where possible, experimental preparation was carried out in a laminar flow bench to eliminate dust contamination. Syringes were used to pump fluids through capillaries for cleaning or preparation of the internal surface. Prior to use, all glassware was sonicated in absolute ethanol and acetone, before being cleaned using Nochromix (Sigma-Aldrich). The glass was then thoroughly rinsed using hot deionised water and dried in a stream of nitrogen gas. Clean, hydrophilic glass was silanized shortly before measurements to produce hydrophobic capillaries. The glass was immersed for ca. 15 minutes in a 3~mM solution of octadecyltrichlorosilane in 80\% bicyclohexyl and 20\% chloroform, then rinsed abundantly and sonicated in chloroform. This procedure was performed in a glove box, with an atmosphere of nitrogen.

Contact angles were measured with a CAM 200 system from KSV Instruments. For all experiments using glass capillaries, a borosilicate glass microscope slide was silanized using the same methodology as the capillary tubes. The mean contact angles measured at two locations on a silanized test slide were $109\pm1$\textdegree\space (advancing) and $88\pm1$\textdegree\space (receding), with hysteresis caused by the roughness of the slide. It was found that the glass capillaries were sometimes imperfectly coated with the silane, as discussed in the Results. The previous hydrophobic treatments used in capillary experiments also used silanization: glass was exposed to dimethyl octyl chlorosilane \cite{828} and propylsilane.\cite{829} In both cases, it is assumed that $\theta_c$ was less than 90\textdegree, because the capillaries were penetrated by a reservoir. The average value of the contact angle of water on PTFE generally lies between $\sim$105 and 115\textdegree ,\cite{793a, 793b} but varies with measurement technique and material impurities. 

Superhydrophobic surfaces were prepared on copper substrates following an electroless deposition methodology.\cite{702} The measured advancing and receding contact angles on these substrates were 173~$\pm$~2\textdegree\space and 161~$\pm$~2\textdegree\space respectively. Because of the high water repellence and the low roll-off angle of these surfaces, droplets typically roll off the surface very easily. However, some droplets become pinned to imperfections on the surface, and can be imaged by optical microscopy. On these imperfections the contact angle is slightly lower than (but close to) that stated above.

\begin{table}
\caption{Details and parameters for the two types of capillary tubing used in experiments. The inner diameter is equal to $2r_t$. Bo is the Bond number, $h_1$ is the meniscus height at which entrance effects become significant and $t_2$ is the time after which viscous effects dominate. Calculations are for water at 20\textdegree C.}
\label{Tab}
\begin{center}
\begin{tabular}{ccccccc}\hline\hline
Material&Supplier&Inner Diameter&Outer Diameter&Bo&$h_1$&$t_2$\\
&&$\mu$m&$\mu$m&x~$10^{-3}$&$\mu$m&ms\\\hline
Borosilicate Glass&Capillary Tube Supplies&100&110&0.34&25&2.5\\
PTFE&Cole-Parmer&305&762&3.12&76&23.1\\
\hline\hline
\end{tabular}
\end{center}
\end{table}

\subsection{Method}

A schematic of the typical experimental output is shown in Fig.~\ref{Fig1b}. A horizontal superhydrophobic substrate was sprayed with a stream of water dispensed from a syringe, producing many droplets with a range of sizes. Due to the superhydrophobic nature of the substrate, the small sessile droplets retain an almost perfectly spherical shape. Capillaries were positioned vertically, with the end of the tube close to the substrate, and aligned so that the centre of the capillary was directly above the highest point of the droplet. The imaging sequence was commenced and the stage smoothly raised to bring the droplet into contact with the end of the tube. Experiments were carried out under ambient laboratory conditions.

Sequences were captured using the CAM 200 system, with a minimum interframe time of 16~ms. Images were spatially calibrated using the known outer tube diameter. When droplet asphericity was observed (often for larger drops), $r_t$ was taken to be the average of horizontal and vertical radii $r_{avge}$. The measurement of $h'$, from the base of the tube to the central height of the meniscus, is an estimate for $h$ (see Fig.~\ref{Fig1b}), because the position at which the meniscus meets the inner edge of the tube is not visible in photographs. 

\section{Results and Discussion}

\subsection{Uptake of Nonwetting Fluids by Dry Capillaries\label{3.1}}

\begin{figure}
\begin{center}
\subfigure[]{\label{Fig2a}\includegraphics[width=5.5cm]{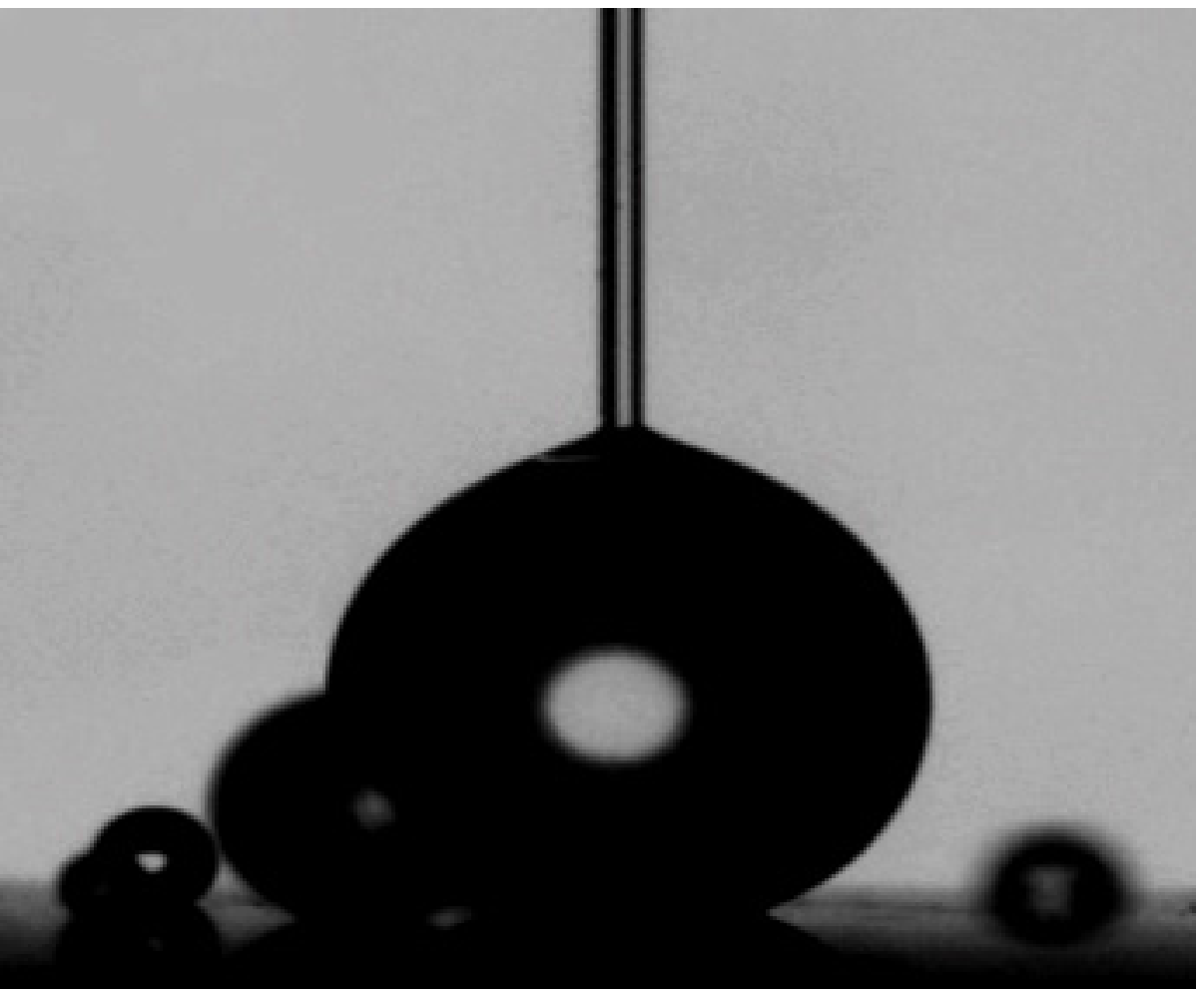}}
\subfigure[]{\label{Fig2b}\includegraphics[width=5.5cm]{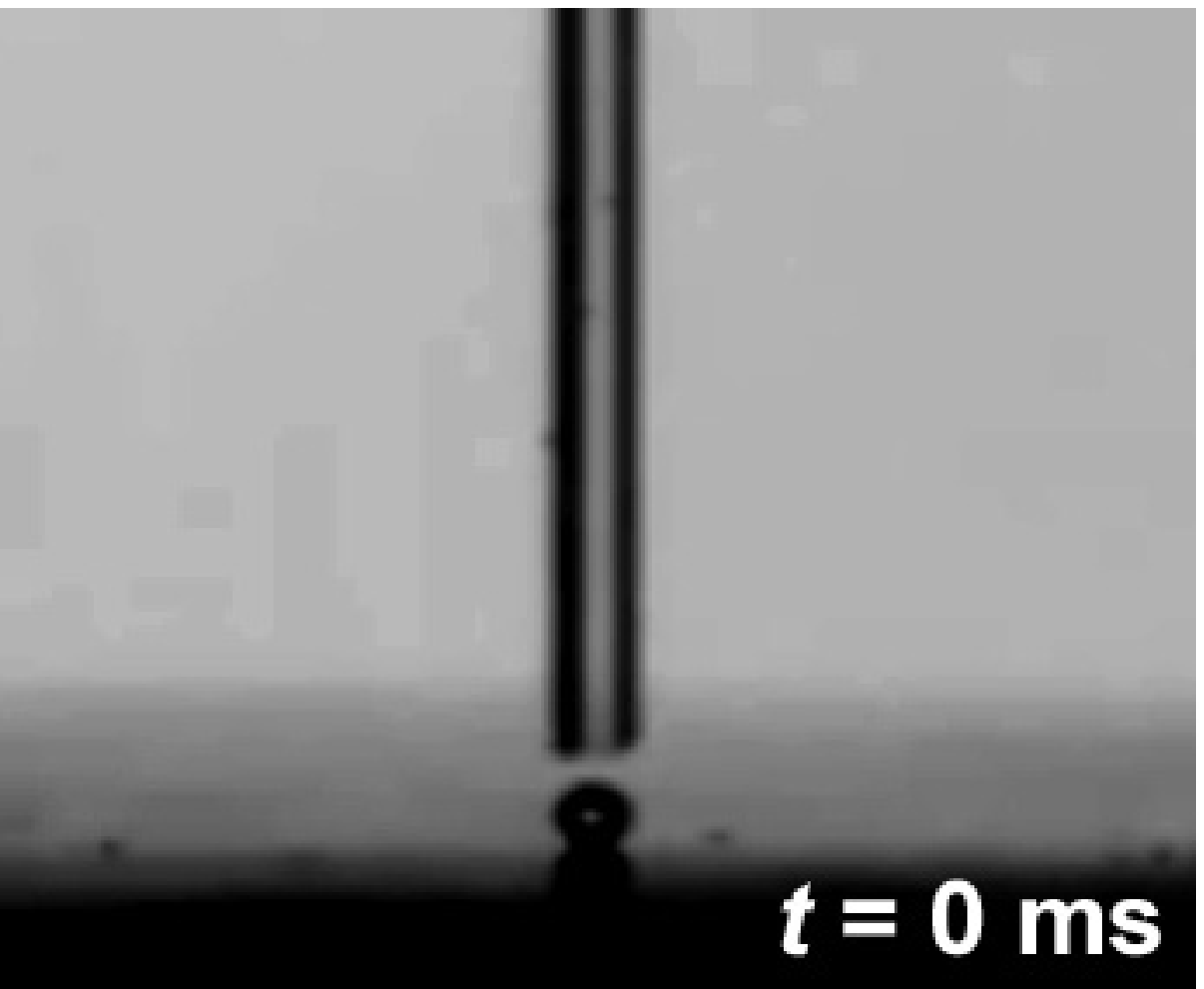}}
\subfigure[]{\label{Fig2c}\includegraphics[width=5.5cm]{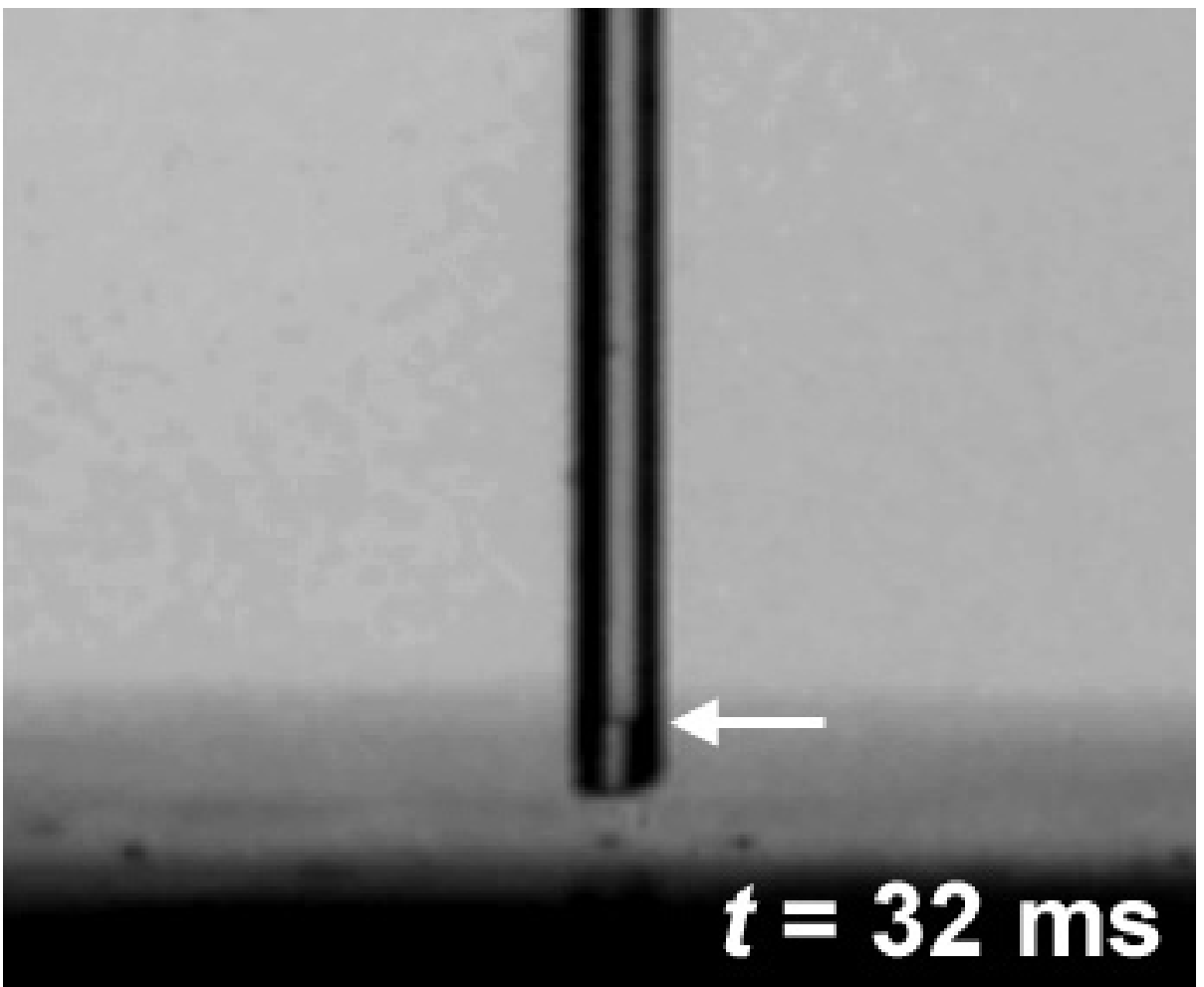}}
\end{center}
\caption{A silane-coated borosilicate glass capillary (advancing $\theta_c\sim$109\textdegree, inner and outer diameters 0.10 and 0.11~mm respectively) is vertically positioned over water droplets on a superhydrophobic surface. No fluid uptake is observed with the capillary in contact with a large water drop (Fig.~\ref{Fig2a}). For a smaller drop (Fig.~\ref{Fig2b}), uptake was observed, with the height of the meniscus indicated by the arrow in Fig.~\ref{Fig2c}.}
\label{Fig2}       
\end{figure}

\begin{figure}
\begin{center}
\subfigure[]{\label{Fig3a}\includegraphics[width=5.5cm]{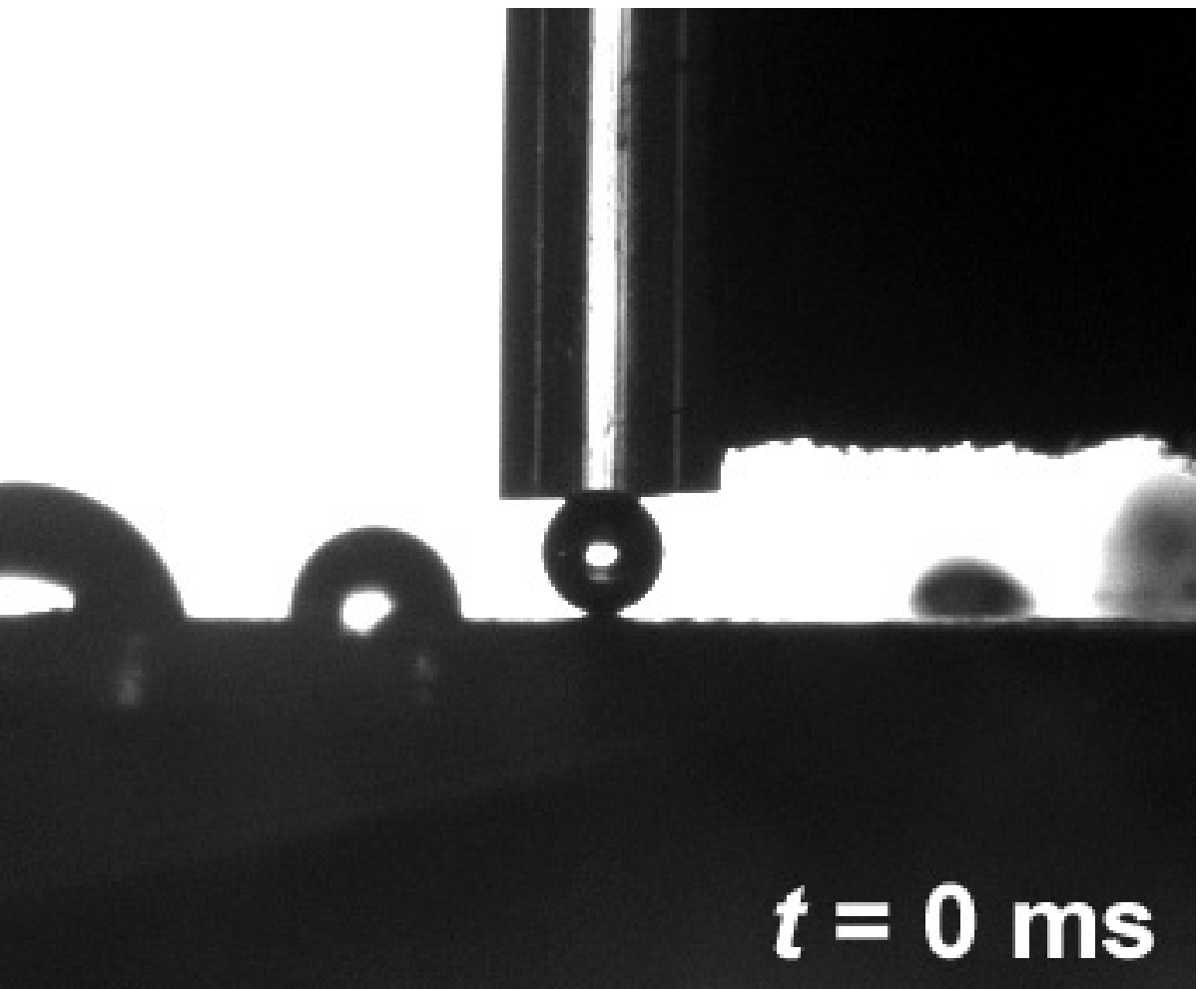}}
\subfigure[]{\label{Fig3b}\includegraphics[width=5.5cm]{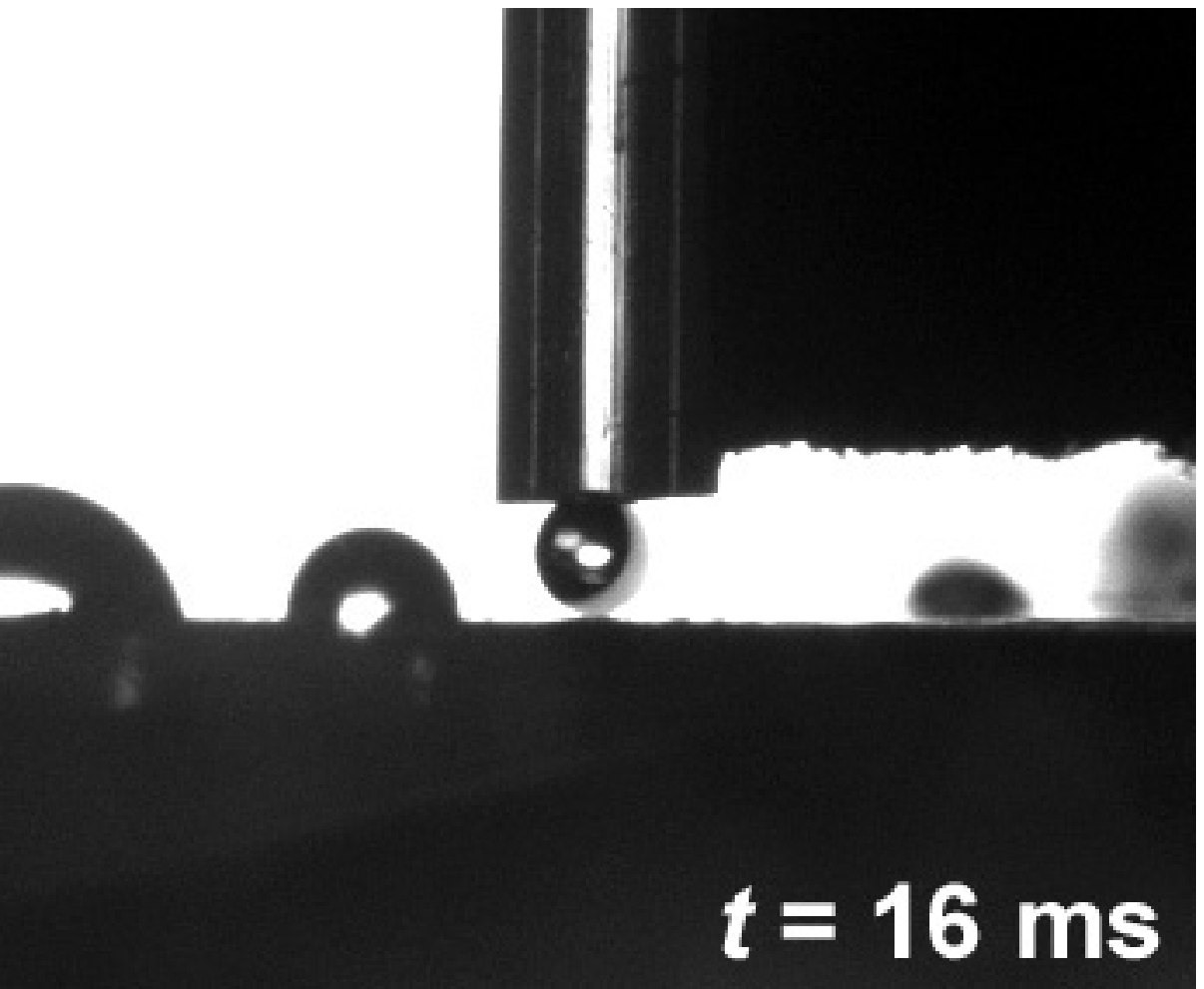}}
\subfigure[]{\label{Fig3c}\includegraphics[width=5.5cm]{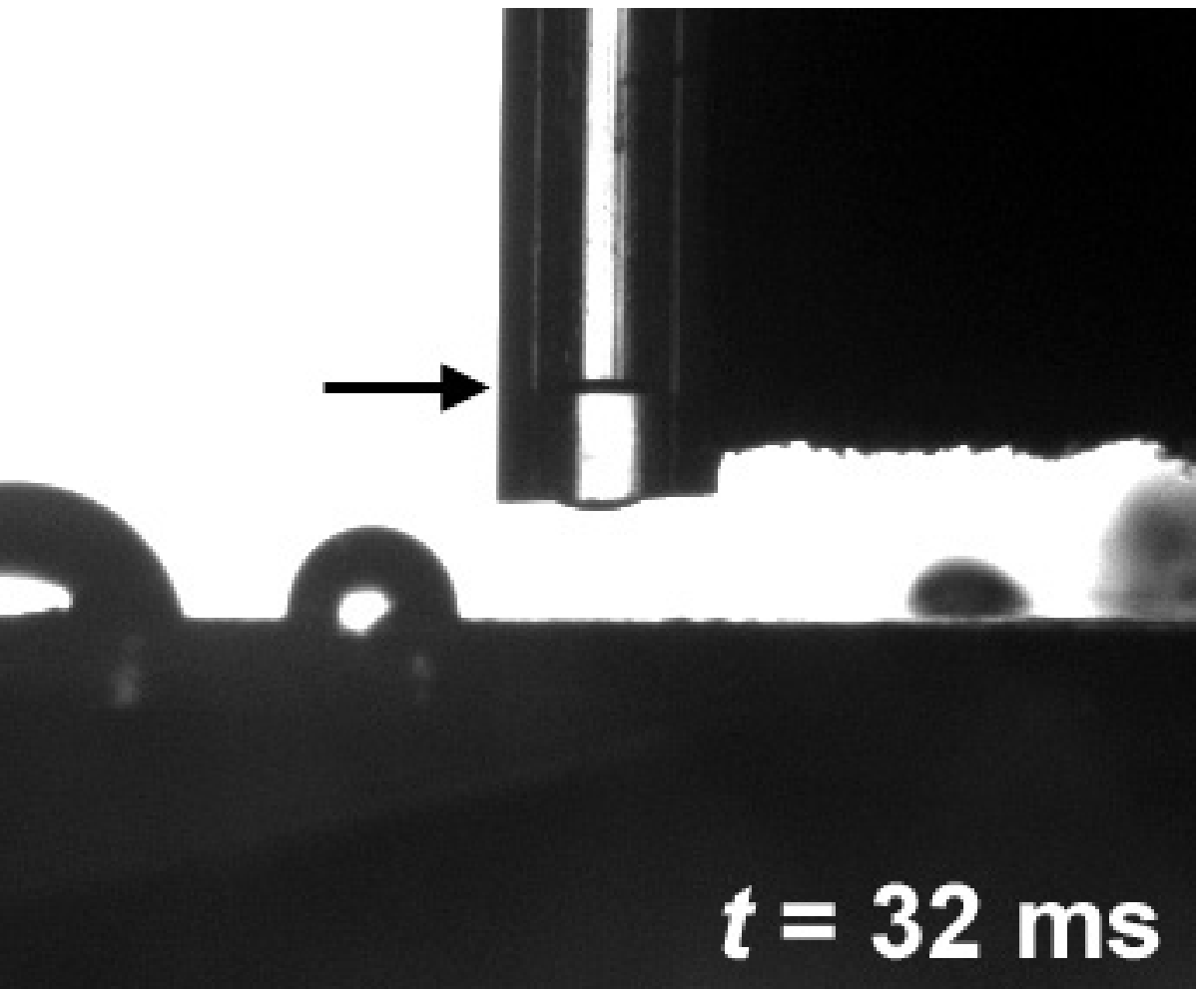}}
\end{center}
\caption{A PTFE capillary (inner and outer diameters 0.31 and 0.76~mm respectively) is brought into contact with a deionised water droplet on a superhydrophobic surface. In Fig.~\ref{Fig3c}, the arrow indicates the height of the meniscus.}
\label{Fig3}       
\end{figure}

Capillary uptake of water drops is demonstrated using silanized borosilicate and PTFE capillaries in Figs.~\ref{Fig2} and \ref{Fig3} respectively. For both types of tube, the capillary could not be penetrated by a large drop or reservoir of water, as shown in Fig.~\ref{Fig2a}. In such cases, the curvature of the surfaces is insufficient to overcome the surface forces between the water and the non-wetting capillaries.

When smaller droplets were brought into contact with the same tubes, uptake of the droplets was observed. Prior to the drop coming into contact with the end of the tube, no fluid is observed in either capillary (Figs.~\ref{Fig2b} and \ref{Fig3a}). In following photographic images, the droplet has disappeared and a meniscus is clearly visible within each tube, indicating that uptake has occurred. Fig.~\ref{Fig3b} captures uptake in progress. Although it is clear that uptake only occurs for smaller droplets, results were not consistent enough to establish a clear drop size threshold for uptake by dry capillaries. 

The relative significance of curvature-induced Laplace pressure compared with other mechanisms can be studied by separating the three regimes developed for reservoir experiments and discussed in the Introduction. To distinguish the regimes, we use a height $h_1$ to describe the limit for the entrance effects regime, and a time $t_2$ to determine when viscous effects should dominate, marking the transition between the Bosanquet and Lucas-Washburn regimes. We use a value of $h_1$ equal to $r_t$~/~2, consistent with values derived for wetting capillaries ($\theta_c=0$\textdegree).\cite{824,809} The characteristic viscous time scale $t_2$ is equal to $\rho r_t^2$~/~$\eta$. Fries and Dreyer \cite{817} define an `inertial-viscous' regime, bounded by low ($t_2$~/~4 \cite{809} or $t_2$~/~16 \cite{824}) and high ($\sim2.1t_2$ \cite{817}) calculations of this transition time.

Referring to Table~\ref{Tab}, we find that the penetration events extend to all three regimes. Entrance effects are obviously important for the initial wetting of the tube, but in both Figs.~\ref{Fig2} and \ref{Fig3}, the capillaries are filled past $h_1$ and the droplet is entirely absorbed. The value of $t_2$ is much less than the time between sequential photographs for the silanized glass capillaries, but approximately equal to 1.5 interframe times (24~ms) for PTFE. Therefore, the dynamics are dominated by entrance effects, inertia and viscosity at different stages during the observed uptake. The low Bond numbers indicate that gravity should not be significant for the uptake process within the tube. Note that gravity significantly affects the shape of relatively large droplets, such as those used in some of the experiments described below. 

In both Figs.~\ref{Fig2} and \ref{Fig3}, the drops have more affinity for the capillary than the substrate, and jump into full contact with the tube - the drop may even avoid simultaneous contact with the capillary and substrate. In other experiments, and especially for larger droplets (e.g. Fig.~\ref{Fig4}) the drop remains in contact with the substrate, either because of gravity or the affinity of the drop for an imperfection on the superhydrophobic surface. Adherence to the surface could significantly affect drop pressure, and would cause different dynamic behaviour to the `jump into contact' case.

\subsection{Drop Size Dependence of Water Motion in PTFE Capillaries\label{3.2}}

A series of experiments was carried out with some fluid pre-loaded into PTFE capillaries. Using the balance of forces for uptake (Eq.~\ref{eq:812} or \ref{eq:S&H}), we can predict that the initial direction of meniscus motion will depend on whether the initial drop radius is greater or less than $r_{max}$.  

\begin{figure}
\begin{center}
\subfigure[]{\label{Fig4a}\includegraphics[width=4.25cm]{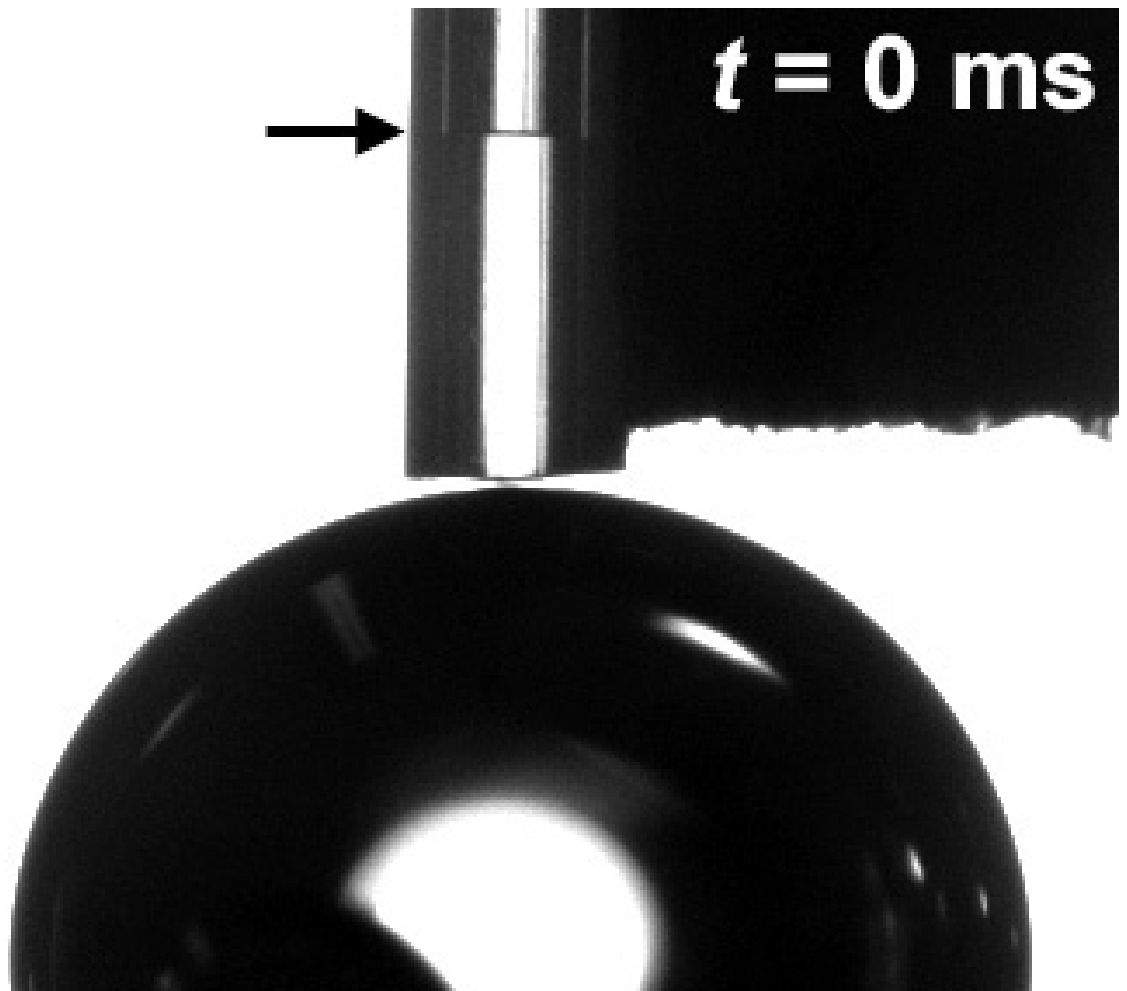}}
\subfigure[]{\label{Fig4b}\includegraphics[width=4.25cm]{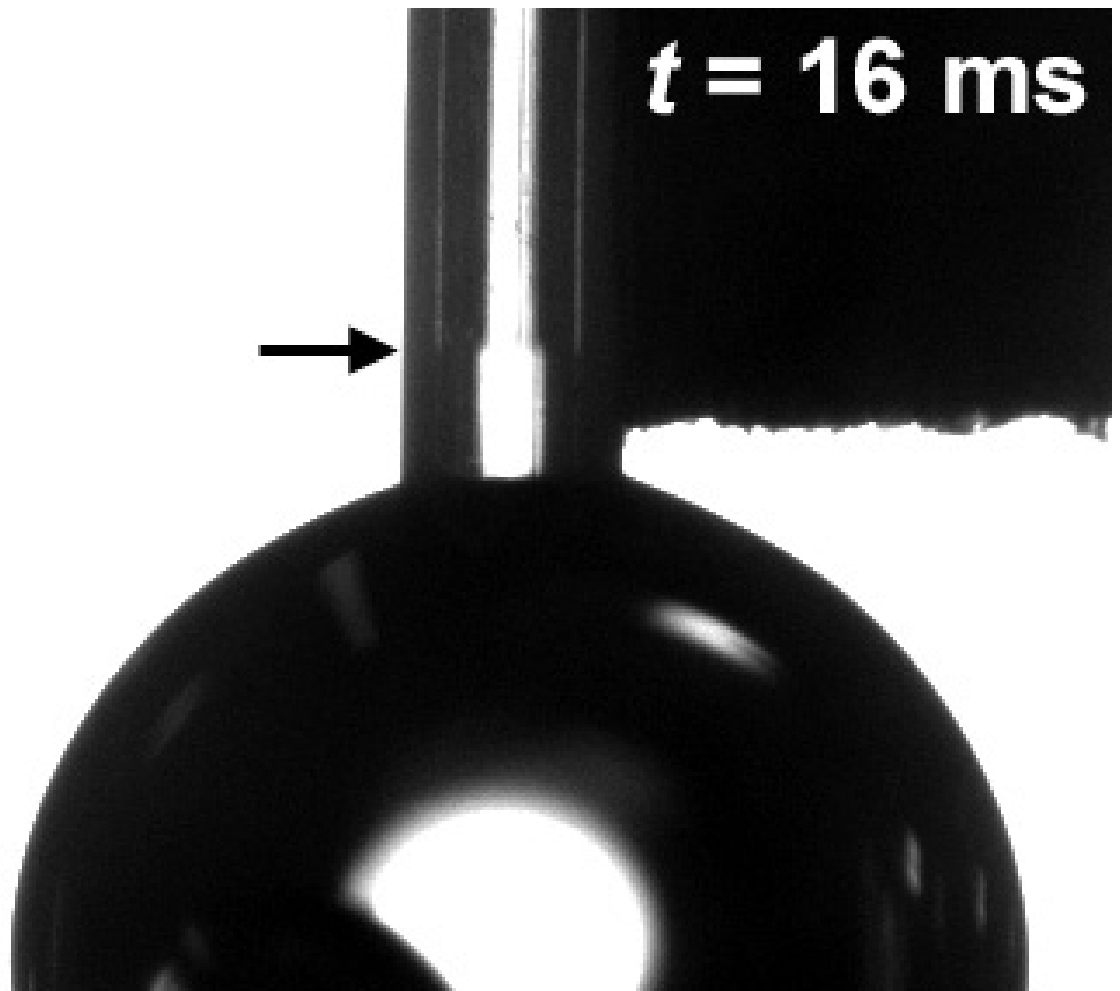}}
\subfigure[]{\label{Fig4c}\includegraphics[width=4.25cm]{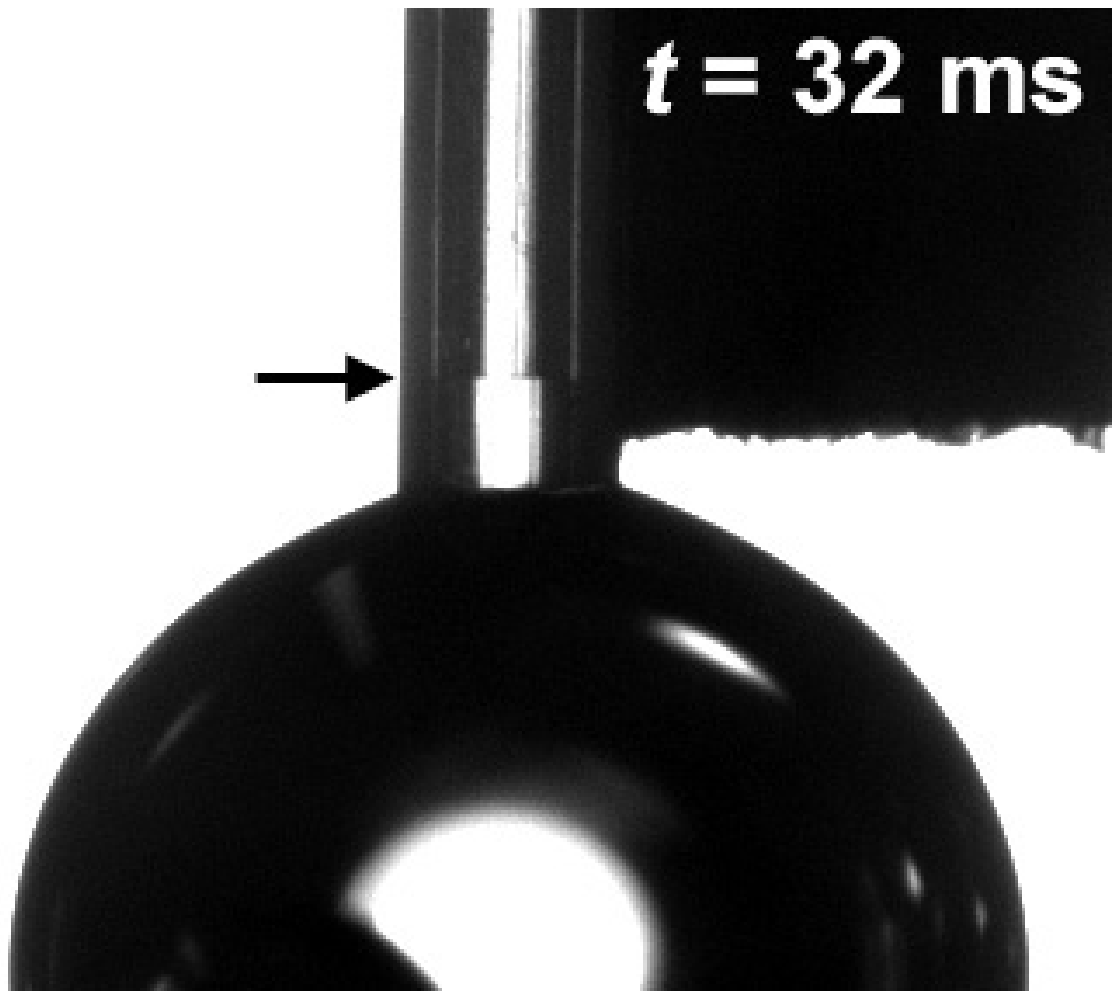}}
\subfigure[]{\label{Fig4d}\includegraphics[width=4.25cm]{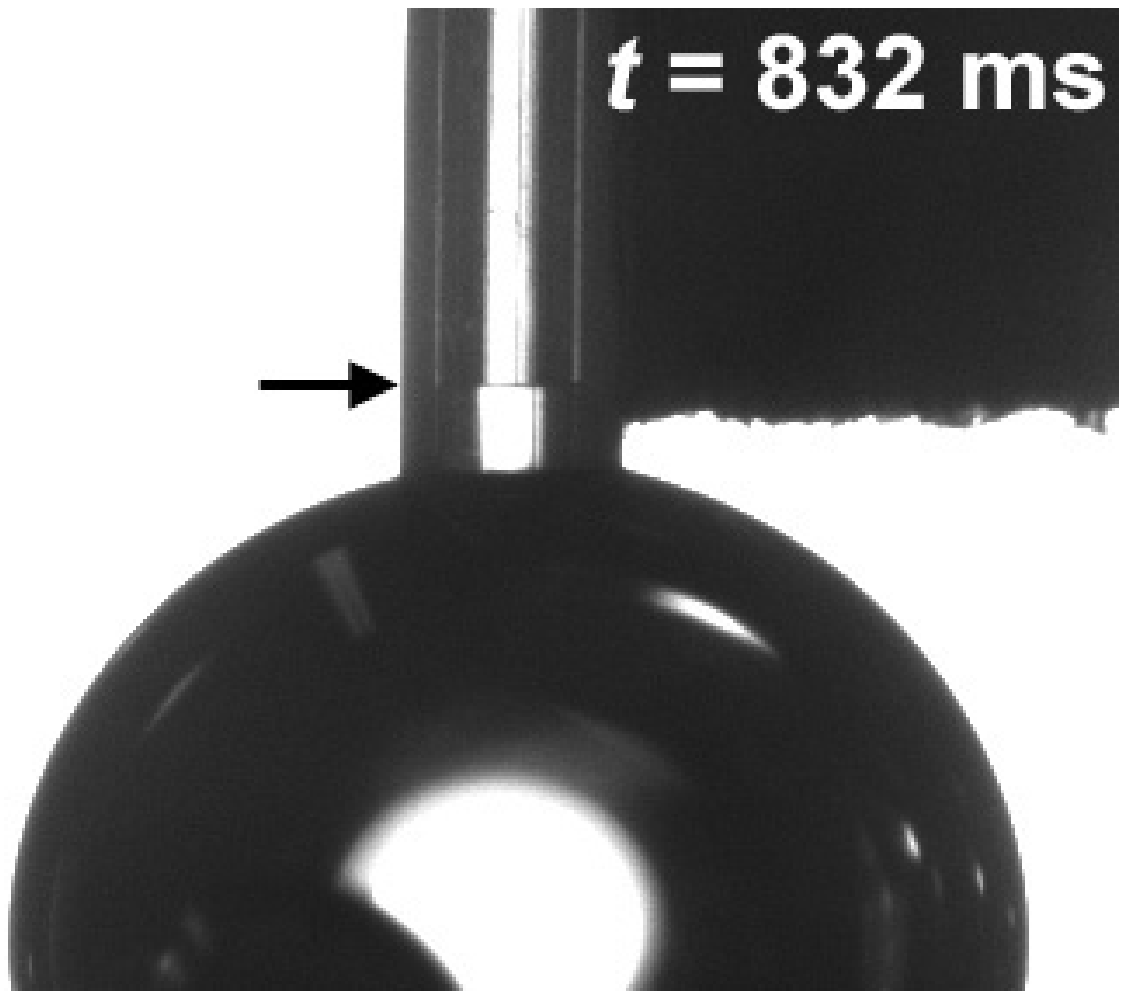}}
\end{center}
\caption{A partially-filled PTFE capillary (inner and outer diameters 0.31 and 0.76~mm respectively) is brought into contact with a relatively large droplet (width 3.4~mm). The meniscus moves downwards, appearing blurred (in motion) in Fig~\ref{Fig4b}. The position of the capillary support relative to the droplet indicates that the capillary is stationary during meniscus motion.}
\label{Fig4}       
\end{figure}

A step change in the meniscus height was typically observed in experiments with pre-filled capillaries, and the direction of meniscus motion indeed varied with drop size. Small drops caused an increase in the meniscus height, while larger drops could extract fluid from a capillary. Fig.~\ref{Fig4} demonstrates a typical fluid drainage result, in which $h'$ decreased rapidly initially, but motion was practically arrested within 100~ms. A reduced meniscus level is observed over the first two sequential frames (Figs.~\ref{Fig4a} to \ref{Fig4c}), but the meniscus is nearly stationary thereafter (Fig.~\ref{Fig4d}). 

The mechanism which causes the meniscus to stop moving is not clear. In the case of small drops, uptake can be halted by the finite size of the drops themselves. For larger drops, the force balance in the Lucas-Washburn regime (Eq.~\ref{eq:S&H}) predicts that the meniscus would move continuously, completely draining the capillary within the time separating two photographic frames. However, as the meniscus height approaches zero, the increased effect of inertia and the capillary entrance become important. Entrance flow characteristics for fluid drainage are those of a jet, as opposed to a sink for capillary uptake.\cite{814}

A number of dynamic effects relevant to this experiment present opportunities for further study. For a reasonably small drop, changes in drop curvature and shape (and therefore Laplace pressure) should affect the meniscus dynamics. The initial filled tube height and interactions between the drop and the substrate could play a significant role in the motion of the meniscus. The present system also allows the dynamic contact angle of a nonwetting system to be studied. 

Further experiments were carried out in which a large droplet was initially in contact with a partially filled tube, with the meniscus stationary. As the substrate was retracted, the drop remained on the substrate, eventually detaching from the capillary. These drops satisfy $r > r_{max}$, so we predict that the meniscus should move upwards when the drop is removed. Again, observations were consistent with this prediction. Note that, in this case, a complicated geometry emerges as fluid forms a neck between the drop and the capillary immediately prior to the drop disengaging from the capillary.

\begin{figure}
\begin{center}
\includegraphics[width=8.25cm]{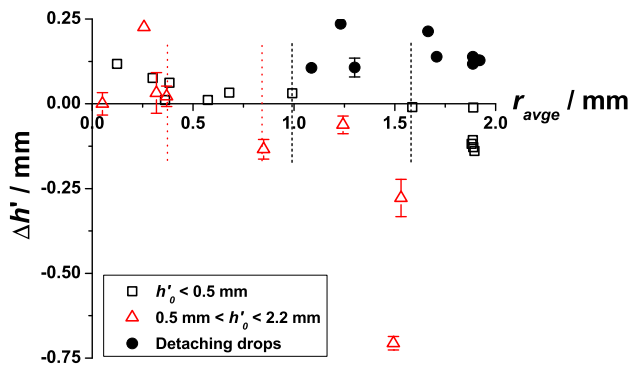}
\end{center}
\caption{The difference between final and initial meniscus heights in PTFE capillaries is plotted as a function of drop radius. The value of $r_{max}$ (at which $\Delta h'=0$) is bounded by vertical lines - black dashes for $h'_0<0.5$ and red dots for $h'_0>0.5$. For the detaching drop data, $h'_0$ ranged between 0.14 and 0.95~mm. Error bars are determined by the spatial resolution of measurements. All horizontal errors (not plotted) are less than $\pm$~0.04~mm; vertical bars are displayed when they are greater than $\pm$~0.02~mm.}
\label{Fig5}       
\end{figure}

Quantitative results for many experiments using PTFE capillaries are summarized in Fig.~\ref{Fig5}. $\Delta h'$ is the difference between the final capillary height and the initial height $h'_0$. The data for detaching drops all have positive values of $\Delta h'$, consistent with the trend described above. The remaining data show a clear qualitative progression from positive $\Delta h'$ for smaller drops to negative values for larger drops, regardless of $h'_0$. The experimental value of $r_{max}$, at which there is zero meniscus motion, appears to be greater when $h'_0$ is smaller. Data for which $h'_0<0.5$~mm indicate a threshold drop radius between 0.99 and 1.59~mm. The equivalent range of $\theta_c$, calculated using Eq.~\ref{eq:rmax}, is between 96\textdegree\space and 99\textdegree. When $h'_0>0.5$~mm, 0.37$<$~$r_{max}$~$<$0.85, and $\theta_c$ lies between 100\textdegree\space and 114\textdegree. These derived contact angles are in reasonable agreement with the expected (static) contact angle for water on PTFE. 

\subsection{Drop Size Dependence of Uptake Speed}

In one particular experimental configuration, the extended duration and slow speed of uptake allowed us to clearly observe the dependence of uptake speed on drop size. These experiments were carried out using a particular glass capillary which had been silanized, but was nevertheless able to be wet by a reservoir of water. The effective static contact angle was therefore less than 90\textdegree , although the glass was still relatively hydrophobic, because capillary uptake was slow. In this case, the silane appears to have formed an imperfect monolayer at the end and on the interior of the tube. 

\begin{figure}
\begin{center}
\includegraphics[width=8.25cm]{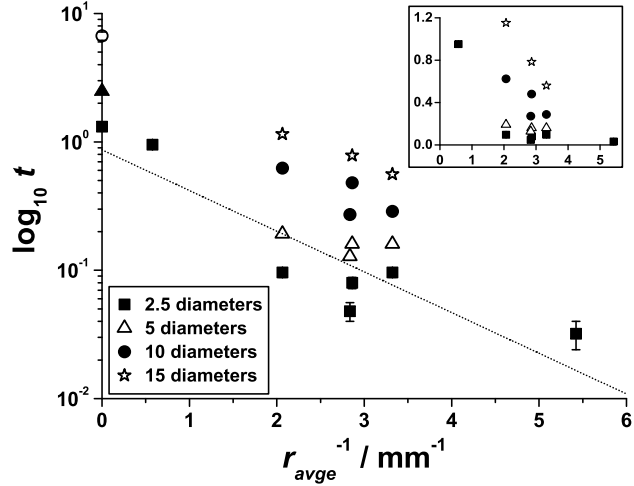}
\end{center}
\caption{Capillary uptake rates for seven experiments, including two at $r_{avge}^{-1}\approx2.8$~mm$^{-1}$, using an imperfectly silanized glass capillary ($r_t=50$~$\mu$m). For each experiment, the log of the time (in seconds) taken for the meniscus to travel a certain distance up the tube (measured in tube diameters, see legend) is plotted against the inverse average drop size. Inset, the data for finite $r_{avge}$ is replotted with a linear vertical scale. In some experiments, data was not recorded for 5, 10 or 15 diameters because the upper part of the tube was not in the camera's field of view. The dotted line is a line of best fit through the data for 2.5 diameters. Error bars indicate time resolution of the photography.}
\label{FigDrops}     
\end{figure}

Fig.~\ref{FigDrops} shows the time taken for the meniscus to reach a certain height ($h'$) when the capillary tube was filled from an initially dry state. The data are qualitatively consistent with theory predicting that uptake speed increases for smaller drops.\cite{790,757,758} The hydrophobic nature of the glass is demonstrated by slow meniscus motion in experiments when compared with predicted speeds for a wetting capillary. For example, the meniscus travels 10 diameters ($h=$1~mm) in $\sim$3~s when the capillary contacts a reservoir, whereas the equivalent time for a wetting tube is less than 1~ms using the Lucas-Washburn approach (Eq.~\ref{eq:L&W}). We have also observed uptake from water reservoirs by clean glass capillaries of inner diameter 100~$\mu$m, in which $h'$ reaches 1~mm within 32~ms of initial wetting, at most. The experiments using a reservoir of water encapsulate all three regimes described in the Introduction. 

We can compare these data with previous experiments in which droplet capillary uptake data was obtained by measuring the droplet volume as a function of time.\cite{807} The results of that study showed a small but clear decrease of the experimental wetting rate as the initial water drop radius was increased from 0.6 to 1.0~mm, using glass capillaries of 0.3~mm radius. The associated analysis recognised Marmur's theory,\cite{790} and the comparison with experiments was conducted relative to the $t=0$ limit ($h$~$\propto$~$t$ regime). The authors did not consider drop size dependence to be significant relative to other effects in their analysis.

Scatter observed in Fig.~\ref{FigDrops} can be partly attributed to imperfect silanization of the capillary. However, random errors are also apparent in the results of Sections~\ref{3.1} and \ref{3.2}. To address these random errors, the key variables to control in future investigations are precise alignment of the droplet with the tube, the velocity of substrate motion and the `jump into contact', the forces acting on droplets while they are outside the tube (including surface interactions and gravity), and atmospheric conditions.

Imperfections in the section of the capillary end may significantly affect the dynamics, but would lead to systematic rather than random variations. Likewise, issues relating to the theoretical analysis are not responsible for random experimental errors. Further work should address some outstanding theoretical issues, which prevent meaningful comparison between the time-dependent data and existing theory (see Introduction). The dynamic contact angle could be studied for systems with a high static contact angle. `End effects' should be studied for finite-sized drops, partially-filled capillaries and fluid retraction rather than penetration. The relative effects of inertia and viscous drag should also be investigated for these systems, in a manner similar to existing work for reservoirs and empty, wetting tubes. Finally, the role of gravity becomes important in this process (even for thin tubes) as the capillary number approaches 1. This was demonstrated in the present work when large droplets preferentially remained on a substrate rather than adhering to a capillary for which they had more affinity. 

\section{Conclusion}

In this paper we have shown that small droplets can penetrate capillaries with static contact angles greater than 90\textdegree, challenging the textbook assumption that liquids cannot spontaneously penetrate non-wetting capillaries. We have experimentally demonstrated the influence of drop size on fluid uptake by single cylindrical capillaries. The direction of meniscus motion in a PTFE tube depends on drop size; the threshold drop radius gives a quantitative prediction for the contact angle of the inner surface of the capillary which is broadly consistent with literature values. The rate of liquid uptake by a silanized (but wetting) capillary is greater for smaller droplets. These phenomena are conceptually simple, being attributable to the Laplace pressure within a small droplet. They are qualitatively consistent with theoretical predictions, and broadly applicable in nano- and microfluidics. For example, this work introduces the possibility of using non-wetting capillaries as micro-pumps, to pick-up, transport, and deposit nano- and micro-droplets with little loss. Also, this work could significantly advance our understanding of CNT growth from metal catalyst particles and suggest new methods for fabricating composite metal-CNT materials. The driving mechanism extends beyond the usual Lucas-Washburn approach, affecting regimes in which inertial and entrance effects are significant. There are numerous experimental variables to be probed and theoretical developments to be made in future work. 

\section{Acknowledgements}

This collaboration was supported by the Royal Society of New Zealand's International Science and Technology Linkages Fund and Marsden Fund. The authors would like to thank other members of the IRL nano- and microfluidics team for practical assistance and useful discussions.

\end{document}